\documentclass[10pt]{revtex4}
\usepackage{amssymb}  
\usepackage{latexsym}
\usepackage{epsfig}
\begin{document}

\title{Quintessence Ghost Dark Energy Model}

\author{Ahamd Sheykhi$^{1,2}$\footnote{sheykhi@mail.uk.ac.ir} and Ali Bagheri$^{1}$ }

\address{$^1$Department of Physics, Shahid Bahonar University, P.O. Box 76175, Kerman, Iran\\
         $^2$Research Institute for Astronomy and Astrophysics of Maragha (RIAAM), Maragha,
         Iran}

\begin{abstract}
A so called `` ghost dark energy" was recently proposed to explain
the present acceleration of the universe expansion. The  energy
density of ghost dark energy, which originates from Veneziano ghost
of QCD, is proportional to the Hubble parameter, $\rho_D=\alpha H$,
where $\alpha$ is a constant which is related to the QCD mass scale.
In this paper, we establish the correspondence between ghost dark
energy and quintessence scalar field energy density. This connection
allows us to reconstruct the potential and the dynamics of the
quintessence scalar field according to the evolution of ghost energy
density.

\end{abstract}

 \maketitle

\section{Introduction}
A wide range of cosmological observations, direct and indirect,
provide an impressive evidence in favor of the present acceleration
of the cosmic expansion. To explain this acceleration, in the
context of standard cosmology, we need an anti gravity fluid with
negative pressure, usually dubbed ``dark energy" in the literature.
The first and simple candidate for dark energy is the cosmological
constant with equation of state parameter $w=-1$ which is located at
the central position among dark energy models both in theoretical
investigation and in data analysis \cite{wein}. However,  there are
several difficulties with cosmological constant.  For example, it
suffers the so-called {fine-tuning} and {cosmic coincidence}
problems. Besides, the origin of it is still a much source of doubt.
Furthermore, the accurate data analysis, show that the time varying
dark energy gives a better fit than a cosmological constant and in
particular, $w$ can cross $-1$ around $z=0.2$ from above to below
\cite{Alam}. Although the galaxy cluster gas mass fraction data do
not support the time-varying $w$ \cite{chen}, an overwhelming flood
of papers has appeared which attempt to understand  the $w=-1$
crossing. Among them are a negative kinetic scalar field and a
normal scalar field \cite{Feng}, or a single scalar field model
\cite{MZ}, interacting holographic \cite{Wang1} and interacting
agegraphic \cite{Wei} dark energy models. Other studies on the
$w=-1$ crossing \cite{Noj} and dark energy models have been carried
out in \cite{nojiri}. For a recent review on dark energy models see
\cite{cop}. It is worthy to note that in most of these dark energy
models, the accelerated expansion are explained by introducing new
degree(s) of freedom or by modifying the underlying theory of
gravity.

Recently a very interesting suggestion on the origin of a dark
energy is made, without introducing new degrees of freedom beyond
what are already known, with the dark energy of just the right
magnitude to give the observed expansion \cite{Urban,Ohta}. In this
proposal, it is claimed that the cosmological constant arises from
the contribution of the ghost fields which are supposed to be
present in the low-energy effective theory of QCD
\cite{Wit,Ven,Ros,Na,Kaw}. The ghosts are required to exist for the
resolution of the $U(1)$ problem, but are completely decoupled from
the physical sector \cite{Kaw}. The above claim is that the ghosts
are decoupled from the physical states and make no contribution in
the flat Minkowski space, but once they are in the curved space or
time-dependent background, the cancelation of their contribution to
the vacuum energy is off-set, leaving a small energy density
$\rho\sim H \Lambda^3_{QCD}$, where $H$ is the Hubble parameter and
$\Lambda_{QCD}$ is the QCD mass scale of order a $100 MeV$. With
$H\sim 10^{-33} eV$, this gives the right magnitude $\sim (3\times
10^{-3} eV)^4$ for the observed dark energy density. This numerical
coincidence is remarkable and also means that this model gets rid of
fine tuning problem \cite{Urban,Ohta}. The advantages of this new
model compared to other dark energy models is that it is totally
embedded in standard model and general relativity, one needs not to
introduce any new parameter, new degree of freedom or to modify
gravity. The dynamical behavior of the ghost dark energy (GDE) model
in flat \cite{CaiGhost} and non flat \cite{shmov} universe have been
studied in ample details.

On the other side, the scalar field model can be regarded as  an
effective description of an underlying dark energy theory. Scalar
fields naturally arise in particle physics including supersymmetric
field theories and string/M theory. Therefore, scalar field is
expected to reveal the dynamical mechanism and the nature of dark
energy. However, although fundamental theories such as string/M
theory do provide a number of possible candidates for scalar fields,
they do not predict its potential $V(\phi)$ uniquely. Consequently,
it is meaningful to reconstruct the potential $V(\phi)$ from some
dark energy models possessing some significant features of the
quantum gravity theory, such as holographic and agegraphic dark
energy models. In the framework of holographic and agegraphic dark
energy models, the studies on the reconstruction of the quintessence
potential $V(\phi)$  have been carried out in \cite{Zhang} and
\cite{ageQ}, respectively. Till now, quintessence reconstruction of
ghost energy density has not been done.

In this paper we are interested in that if we assume the GDE
scenario as the underlying theory of dark energy, how the low-energy
effective scalar-field model can be used to describe it. In this
direction, we can establish the correspondence between the GDE and
quintessence scalar field, and describe GDE in this case effectively
by making use of quintessence. We shall reconstruct the quintessence
potential and the dynamics of the scalar field in the light of the
GDE.
\section{Quintessence Ghost dark energy}
We assume the GDE is accommodated in a flat
Friedmann-Robertson-Walker (FRW) which its dynamics is governed by
the Friedmann equation
\begin{eqnarray}\label{Fried}
H^2=\frac{1}{3M_p^2} \left( \rho_m+\rho_D \right),
\end{eqnarray}
where $\rho_m$ and $\rho_D$ are the energy densities of pressureless
matter and GDE, respectively. We define the dimensionless density
parameters as
\begin{equation}
\Omega_m=\frac{\rho_m}{\rho_{\rm cr}},\ \ \
\Omega_D=\frac{\rho_D}{\rho_{\rm cr}},\ \
\end{equation}
where the critical energy density is $\rho_{\rm cr}={3H^2 M_p^2}$.
Thus, the Friedmann equation can be rewritten as
\begin{equation}\label{fridomega}
\Omega_m+\Omega_D=1.
\end{equation}
The conservation equations read
\begin{eqnarray}
\dot\rho_m+3H\rho_m&=&0,\label{consm}\\
\dot\rho_D+3H\rho_D(1+w_D)&=&0\label{consd}.
\end{eqnarray}
The ghost energy density is proportional to the Hubble parameter
\cite{Ohta,CaiGhost}
\begin{equation}\label{GDE}
\rho_D=\alpha H,
\end{equation}
where $\alpha$ is a constant of order $\Lambda_{\rm QCD}^3$ and
$\Lambda_{\rm QCD}\sim 100 MeV$ is QCD mass scale. Taking the time
derivative of relation (\ref{GDE}) and using Friedmann equation
(\ref{Fried}) we find
\begin{equation}\label{dotrho}
\dot{\rho}_D=-\frac{\alpha }{2 M_p^2} \rho_D(1+u+w_D).
\end{equation}
where $u=\rho_m/\rho_D$ is the energy density ratio. Inserting this
relation in continuity equation (\ref{consd}) and using Eq.
(\ref{fridomega}) we find
\begin{equation}\label{wD3}
w_D=-\frac{1}{2-\Omega_D}.
\end{equation}
At the early time  where $\Omega_D\ll 1$ we have $w_D=-1/2$, while
at the late time where $\Omega_D\rightarrow 1$ the GDE mimics a
cosmological constant, namely $w_D= -1$. In figure 1 we have plotted
the evolution of $w_D$ versus scale factor $a$. From this figure we
see that $w_D$ of the GDE model cannot cross the phantom divide and
the universe has a de Sitter phase at late time.
\begin{figure}[htp]
\begin{center}
\includegraphics[width=8cm]{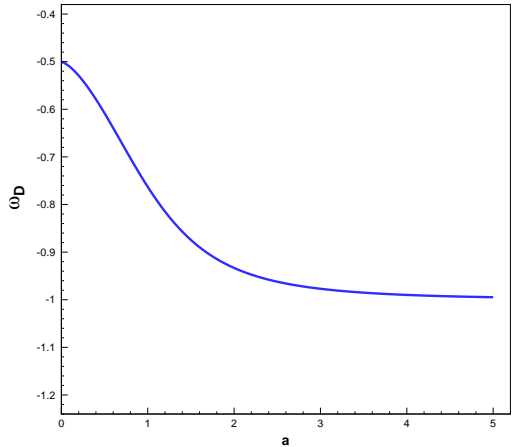}
\caption{The evolution of $w_D$ for GDE. }\label{fig1}
\end{center}
\end{figure}

Now we are in a position to establish the correspondence between GDE
and quintessence scaler field. To do this,
 we assume the quintessence scalar field model of dark energy is
the effective underlying theory. The energy density and pressure
of the quintessence scalar field are given by
\begin{eqnarray}\label{rhophi}
\rho_\phi=\frac{1}{2}\dot{\phi}^2+V(\phi),\\
p_\phi=\frac{1}{2}\dot{\phi}^2-V(\phi). \label{pphi}
\end{eqnarray}
Thus the potential and the kinetic energy term can be written as
\begin{eqnarray}\label{vphi}
&&V(\phi)=\frac{1-w_\phi}{2}\rho_{\phi},\\
&&\dot{\phi}^2=(1+w_\phi)\rho_\phi. \label{ddotphi}
\end{eqnarray}
In order to implement the correspondence between GDE and
quintessence scaler field, we identify $\rho_\phi=\rho_D$ and
$w_\phi=w_D$. Using Eqs.
 (\ref{GDE}) and (\ref{wD3}) as well as relation $\dot{\phi}=H \frac{d\phi}{d\ln a}$ we obtain the scalar potential and the dynamics of scalar field as
\begin{eqnarray}\label{vphi2}
V(\phi)&=&\frac{\alpha^2}{6 M_p^2}\times \frac{3-\Omega_D}{\Omega_D(2-\Omega_D)},\\
\frac{d\phi}{d\ln a} &=&\sqrt{3}M_p \sqrt{\frac{\Omega_D
(1-\Omega_D)}{2-\Omega_D}}.\label{dotphi2}
\end{eqnarray}
Integrating yields
\begin{eqnarray}\label{phi1}
\phi(a)-\phi(a_0)=\sqrt{3}M_p
\int_{a_0}^{a}{\frac{da}{a}\sqrt{\frac{\Omega_D
(1-\Omega_D)}{2-\Omega_D}}},
\end{eqnarray}
\begin{figure}[htp]
\begin{center}
\includegraphics[width=8cm]{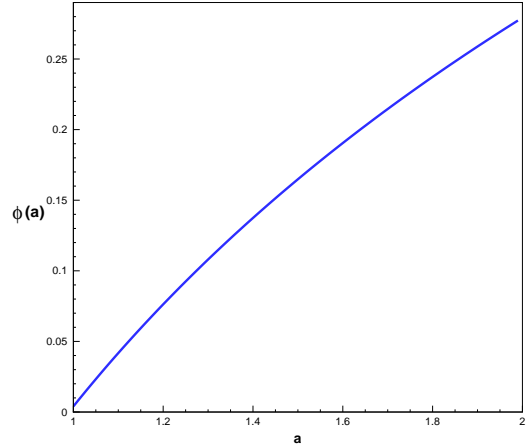}
\caption{The evolution of the scalar field $\phi(a)$ for
quintessence GDE, where $\phi$ is in unit of $\sqrt{3}M_p$.}
\label{fig3}
\end{center}
\end{figure}
\begin{figure}[htp]
\begin{center}
\includegraphics[width=8cm]{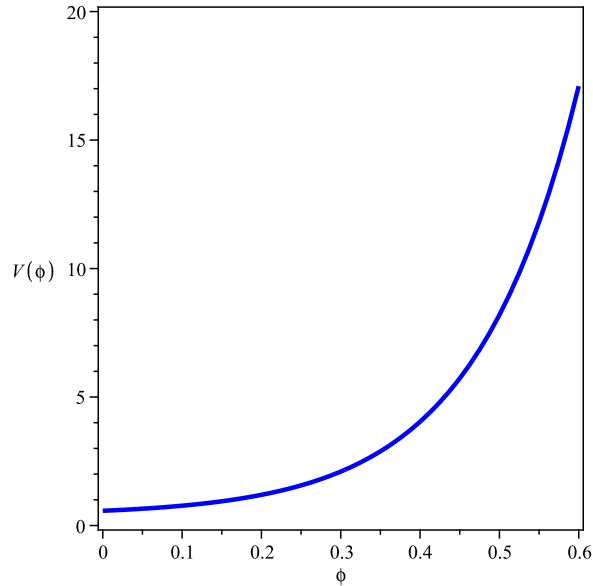}
\caption{The reconstructed potential $V(\phi)$ for quintessence GDE,
where $V(\phi)$ is in unit of $\alpha^2/6M_p^2$. } \label{fig4}
\end{center}
\end{figure}
where we have set $a_0=1$  for the  present value of the scale
factor. The analytical form of the potential in terms of the ghost
quintessence field cannot be determined due to the complexity of the
equations involved. However, we can obtain it numerically. The
reconstructed quintessence potential $V(\phi)$ and the evolutionary
form of the field are plotted in Figs. 2 and 3, where we have taken
$\phi(a_0=1)=0$ for simplicity. From figure 2 we can see the
dynamics of the scalar field explicitly. Obviously, the scalar field
$\phi$ rolls down the potential with the kinetic energy
$\dot{\phi}^2$ gradually decreasing. In other words, the amplitude
of $\phi$ decreases with time in the past.

\section{Interacting Quintessence Ghost dark energy} \label{Int}
Next we generalize our discussion to the interacting case. Although
at this point the interaction may look purely phenomenological but
different Lagrangians have been proposed in support of it (see
\cite{Tsu} and references therein). Besides, in the absence of a
symmetry that forbids the interaction there is nothing, in
principle, against it. In addition, given the unknown nature of both
dark energy and dark matter, which are two major contents of the
universe, one might argue that an entirely independent behavior of
dark energy is very special \cite{wang1,pav1}. Thus, microphysics
seems to allow enough room for the coupling; however, this point is
not fully settled and should be further investigated. The difficulty
lies, among other things, in that the very nature of both dark
energy and dark matter remains unknown whence the detailed form of
the coupling cannot be elucidated at this stage. Since we consider
the interaction between dark matter and dark energy, $\rho_{m}$ and
$\rho_{D}$ do not conserve separately; they must rather enter the
energy balances \cite{pav1}
\begin{eqnarray}
\dot\rho_m+3H\rho_m&=&Q,\label{consm2}\\
\dot\rho_D+3H\rho_D(1+w_D)&=&-Q\label{consd2},
\end{eqnarray}
where $Q$ represents the interaction term and we take it as
\begin{equation}\label{Q}
Q =3b^2 H(\rho_m+\rho_D)=3b^2 H\rho_D(1+u).
\end{equation}
with $b^2$  being a coupling constant. Inserting Eqs. (\ref{dotrho})
and (\ref{Q}) in Eq. (\ref{consd2}) we find
\begin{equation}\label{wD4}
w_D=-\frac{1}{2-\Omega_D}\left(1+\frac{2b^2}{\Omega_D}\right).
\end{equation}
One can easily check that in the late time where
$\Omega_D\rightarrow 1$, the equation of state parameter of
interacting GDE necessary crosses the phantom line, namely,
$w_D=-(1+2b^2)<-1$ independent of the value of coupling constant
$b^2$. For the present time with taking $\Omega_D=0.72$, the phantom
crossing can be achieved provided $b^2>0.1$ which is consistent with
recent observations \cite{wang1}. It is worth mentioning that the
continuity equations (16) and (17) imply that the interaction term
should be a function of a quantity with units of inverse of time (a
first and natural choice can be the Hubble factor $H$) multiplied
with the energy density. Therefore, the interaction term could be in
any of the following forms: (i) $Q\propto H \rho_D$, (ii) $Q\propto
H \rho_m$, or (iii) $Q\propto H (\rho_m+\rho_D)$. We can present the
above three choices in one expression as $Q =\Gamma\rho_D$, where
\begin{eqnarray}
\begin{array}{ll}
\Gamma=3b^2H  \hspace{1.3cm}   {\rm for}\  \  Q\propto H \rho_D, &  \\
\Gamma=3b^2Hu \hspace{1.1cm}   {\rm for} \  \ Q\propto H \rho_m,&  \\
\Gamma=3b^2H(1+u) \ \   {\rm for} \  \ Q\propto H (\rho_m+\rho_D),&
  \end{array}
 \end{eqnarray}
It should be noted that the ideal interaction term must be motivated
from the theory of quantum gravity. In the absence of such a theory,
we rely on pure dimensional basis for choosing an interaction $Q$.
To be more general in this work we choose expression (iii) for the
interaction term. The coupling $b^2$  is taken in the range $[0, 1]$
\cite{HZ}. Note that if $b^2=0$ then it represents the
noninteracting case while $b^2=1$ yields complete transfer of energy
from dark energy to matter ($Q>0$). Although in principle there is
now reason to take $Q>0$ and one may take $Q<0$ which means that
dark matter transfers to dark energy, however, as we will see below
this is not the case. It is easy to show that for $Q<0$, Eq.
(\ref{wD4}) becomes
\begin{equation}\label{wD44}
w_D=-\frac{1}{2-\Omega_D}\left(1-\frac{2b^2}{\Omega_D}\right).
\end{equation}
In the late time where $\Omega_D\rightarrow 1$, we have
$w_D=-(1-2b^2)$, which for $b^2>1/3$ leads to $w_D>-1/3$. This
implies that in the late time where dark energy dominates we have no
acceleration at least for some value of coupling parameter. For the
present time if we take $\Omega_D=0.72$, from Eq. (\ref{wD44}) we
have $w_D=-0.78+2.2 b^2$. Again for $b^2>0.20$ we have $w_D>-1/3$
for the present time. This means that universe is in deceleration
phase at the present time which is ruled out by recent observations.

The behaviour of the equation of state parameter of interacting GDE
is shown in figure 4 for different value of the coupling parameter.
\begin{figure}[htp]
\begin{center}
\includegraphics[width=8cm]{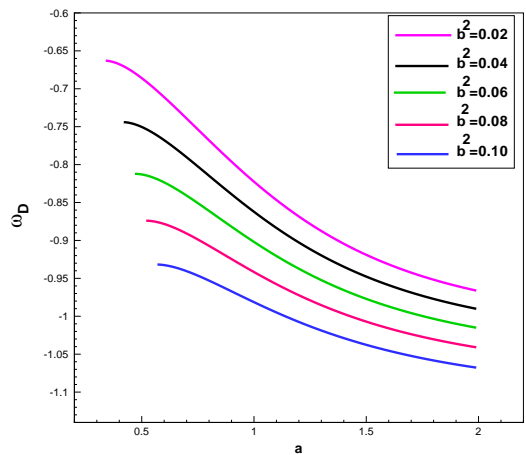}
\caption{The evolution of $w_D$ for interacting GDE. }\label{fig6}
\end{center}
\end{figure}
In the presence of interaction, the evolution of GDE is governed by
the following equation \cite{shmov}
\begin{equation}\label{Omegaprime3}
\frac{d\Omega_D}{d\ln a}=\frac{3}{2} \Omega_D\left[1-\frac{
\Omega_D}{2- \Omega_D}\left(1+\frac{2b^2}{\Omega_D}\right)\right].
\end{equation}
Fig. 5 shows that at the early time $\Omega_D\rightarrow0$ while at
the late time $\Omega_D\rightarrow1$, that is the ghost dark energy
dominates as expected.
\begin{figure}[htp]
\begin{center}
\includegraphics[width=8cm]{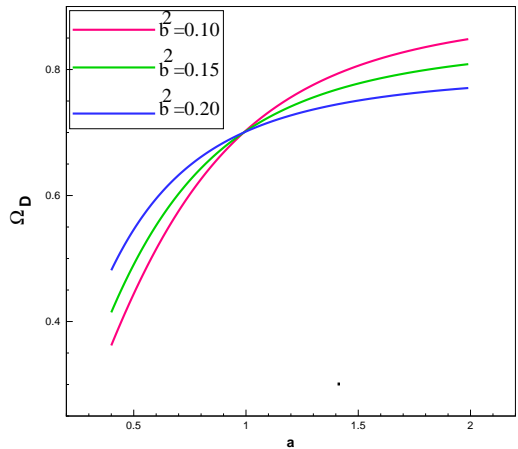}
\caption{The evolution of $\Omega_{D}$ for interacting ghost dark
energy, where we take $\Omega_{D0}=0.72$.}\label{fig7}
\end{center}
\end{figure}
Now we implement a connection between interacting GDE and
quintessence scalar field. In this case the potential and scalar
field are obtained as
\begin{eqnarray}\label{vphi22}
V(\phi)&=&\frac{\alpha^2}{6 M_p^2}\times
\frac{1}{\Omega_D(2-\Omega_D)}
\left(3-\Omega_D+ \frac{2b^2}{\Omega_D}\right),\\
\frac{d\phi}{d\ln a} &=&\sqrt{3}M_p \sqrt{\frac{\Omega_D}
{2-\Omega_D}\left(1-\Omega_D-
\frac{2b^2}{\Omega_D}\right)}.\label{dotphi22}
\end{eqnarray}
Finally we obtain the evolutionary form of the field by
integrating the above equation. The result is
\begin{eqnarray}\label{phi2}
\phi(a)-\phi(a_0)=\sqrt{3}M_p
\int_{a_0}^{a}{\frac{da}{a}\sqrt{\frac{\Omega_D}
{2-\Omega_D}\left(1-\Omega_D- \frac{2b^2}{\Omega_D}\right)}},
\end{eqnarray}
where $\Omega_D$ is now given by Eq. (\ref{Omegaprime3}). The
reconstructed quintessence potential $V(\phi)$ and the evolutionary
form of the field are plotted in Figs. 6 and 7, where again we have
taken $\phi(a_0=1)=0$  for the present time. Selected curves are
plotted for different value of the coupling parameter $b^2$. From
these figures we find out that $\phi$ increases with time while the
potential $V(\phi)$ becomes steeper with increasing $b^2$.
\begin{figure}[htp]
\begin{center}
\includegraphics[width=8cm]{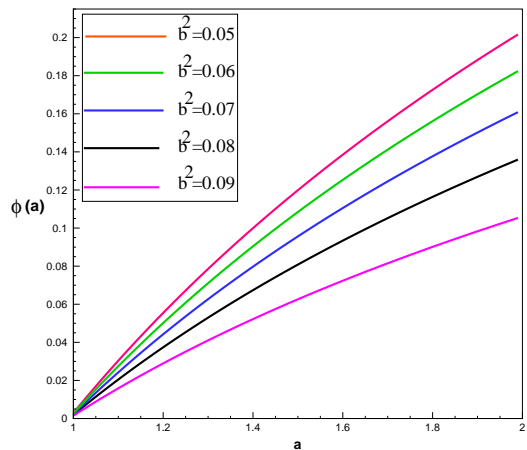}
\caption{The evolutionary form of the scalar field $\phi(a)$ for
interacting quintessence GDE, where $\phi$ is in unit of
$\sqrt{3}M_p$. }\label{fig8}
\end{center}
\end{figure}
\begin{figure}[htp]
\begin{center}
\includegraphics[width=8cm]{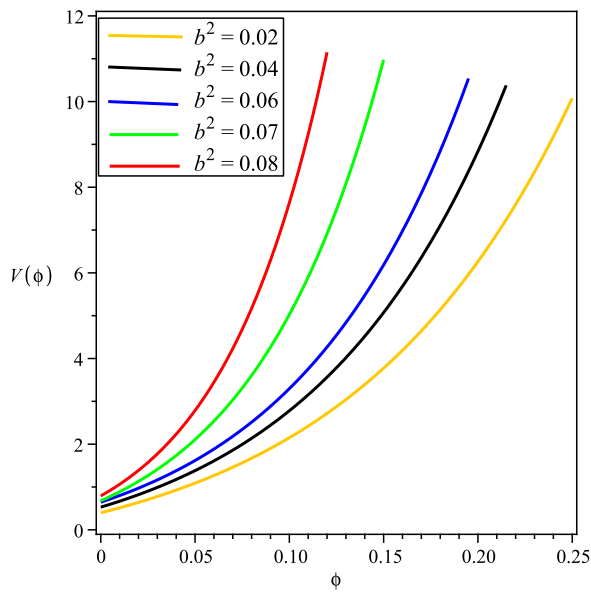}
\caption{The reconstructed potential $V(\phi)$ for interacting
quintessence GDE, where $V(\phi)$ is in unit of $(\alpha^2/6M_p^2)$.
}\label{fig9}
\end{center}
\end{figure}

\section{Conclusion}\label{con}
Considering the quintessence scalar field dark energy model as an
effective description of the underlying theory of dark energy, and
assuming the ghost vacuum energy scenario as pointing in the same
direction, it is interesting to study how the quintessence scalar
field model can be used to describe the ghost energy density. The
quintessence scalar field is specified to an ordinary scalar field
minimally coupled to gravity, namely the canonical scalar field. It
is remarkable that the resulting model with the reconstructed
potential is the unique canonical single-scalar model that can
reproduce the GDE evolution of the universe. In this paper, we
established a connection between the GDE scenario and the
quintessence scalar-field model. The GDE model is a new attempt to
explain the origin of dark energy within the framework of Veneziano
ghost of QCD \cite{Ohta}. If we regard the quintessence scalar-field
model as an effective description of GDE, we should be capable of
using the scalar-field model to mimic the evolving behavior of the
dynamical ghost energy and reconstructing this scalar-field model
according to the evolutionary behavior of GDE. With this strategy,
we reconstructed the potential of the ghost quintessence and the
dynamics of the field according to the evolution of ghost energy
density.

Finally we would like to mention that the aforementioned discussion
in this paper can be easily generalized to other non-canonical
scalar fields, such as K-essence and tachyon. It can also be
extended to the non-flat FRW universe.

\acknowledgments{This work has been supported by Research Institute
for Astronomy and Astrophysics of Maragha, Iran.}

\end{document}